# PrivateXR: Defending Privacy Attacks in Extended Reality Through Explainable AI-Guided Differential Privacy

Ripan Kumar Kundu *
University of Missouri-Columbia

Istiak Ahmed †
University of Missouri-Columbia

Khaza Anuarul Hoque ‡
University of Missouri-Columbia

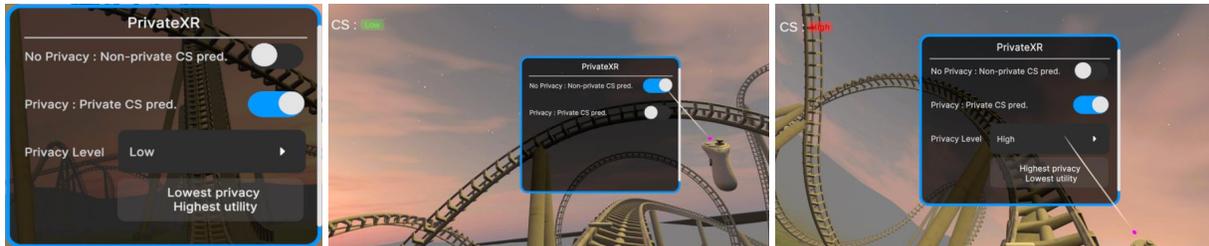

Figure 1: **(a)** The XAI-guided DP-enabled XR UI (PrivateXR) within a virtual roller coaster environment enhances user experience through dynamic privacy controls. **(b)** A player using PrivateXR without DP method (no privacy mode) and model predict low level of cybersickness (CS) severity in real-time, and **(c)** A player using PrivateXR with the XAI-guided DP method selects the high privacy mode, where the model predicts a high level of CS in real-time.

## ABSTRACT

The convergence of artificial intelligence (AI) and extended reality (XR) technologies (AI XR) promises innovative applications across many domains. However, the sensitive nature of data (e.g., eye-tracking) used in these systems raises significant privacy concerns, as adversaries can exploit these data and models to infer and leak personal information through membership inference attacks (MIA) and re-identification (RDA) with a high success rate. Researchers have proposed various techniques to mitigate such privacy attacks, including differential privacy (DP). However, AI XR datasets often contain numerous features, and applying DP uniformly can introduce unnecessary noise to less relevant features, degrade model accuracy, and increase inference time, limiting real-time XR deployment. Motivated by this, we propose a novel framework combining explainable AI (XAI) and DP-enabled privacy-preserving mechanisms to defend against privacy attacks. Specifically, we leverage post-hoc explanations to identify the most influential features in AI XR models and selectively apply DP to those features during inference. We evaluate our XAI-guided DP approach on three state-of-the-art AI XR models and three datasets: cybersickness, emotion, and activity classification. Our results show that the proposed method reduces MIA and RDA success rates by up to 43% and 39%, respectively, for cybersickness tasks while preserving model utility with up to 97% accuracy using Transformer models. Furthermore, it improves inference time by up to $\approx 2\times$ compared to traditional DP approaches. To demonstrate practicality, we deploy the XAI-guided DP AI XR models on an HTC VIVE Pro headset and develop a user interface (UI), namely *PrivateXR*, allowing users to adjust privacy levels (e.g., low, medium, high) while receiving real-time task predictions, protecting user privacy during XR gameplay. Finally, we validate our approach through a user study, which confirms that participants found the PrivateXR UI effective, with satisfactory utility and user experience.

---

*e-mail: rkundu@missouri.edu
†e-mail: ia5qq@missouri.edu
‡e-mail: hoquek@missouri.edu

## 1 INTRODUCTION

Extended reality (XR) data, specifically data related to eye movements, plays a vital role in human cognition, visual attention, and perception [9, 12, 25]. In addition to eye tracking, other eye-related information is commonly utilized in XR with the help of state-of-the-art artificial intelligence (AI)-based techniques, such as machine learning (ML) and deep learning (DL) algorithms for iris recognition [33], gaze prediction [29], and emotion recognition [68]. However, integrating DL/ML for these tasks in XR poses a significant privacy risk [18, 19, 44, 67]. This is because eye-tracking data presents a critical risk to users' privacy, as it captures sensitive attributes about users, such as personal identities, gender, and sexual preferences, based on where they look in the virtual environment. Furthermore, this also introduces the potential risk of re-identification through the captured data by applying re-identification attacks (RDA) in which users' identities are revealed by linking anonymized data to their real-world identity using auxiliary information (e.g., behavioral patterns) through the captured data [19, 54].

*Limitation of prior works.* Researchers have applied differential privacy (DP) mechanisms to different representations of gaze data, such as time series of gaze positions [38], eye images [62], etc, to mitigate privacy risks in XR and provide formal privacy guarantees. In addition, other formal privacy mechanisms for mitigating privacy risks, such as k-anonymity and plausible deniability, are also applied to gaze samples to provide privacy in activity recognition and gaze prediction tasks. Unfortunately, as noted by recent work [11, 19, 20, 39], most prior works applied privacy mechanisms in the eye-tracking sample levels to make the dataset private while ignoring the ML/DL part. Moreover, eye-tracking data are already noisy sources of information; thus, adding additional noise to preserve privacy may ruin the user experience. Indeed, protecting the AI model's privacy is crucial to ensure that individual contributions remain confidential since making only the data private might not guarantee robust protection against privacy breaches [12]. For instance, ML/DL methods are also vulnerable to other types of privacy attacks, such as membership inference attacks (MIA) [65], in which adversaries attempt to determine whether specific individuals' data was used to train a model. Very recently, Kundu et.al. [35] applied MIA against DL-based cybersickness classification models to leak information about individual users in the training data of a VR cybersickness dataset. However, their methods are typically application-specific, focusing only on tasks (i.e., cybersickness), limiting their generalizability to other XR applications. Furthermore, their work only shows the impact of MIA in cybersickness classification tasks. While MIA exposes vulnerabilities in the AI model by determining whether specific user data was used during training, it does not account for de-anonymization (e.g., through re-identification) risks in shared or anonymized datasets, which is common in XR research. Thus, a comprehensive XR privacy risk assessment is required to evaluate, analyze, and mitigate different types of privacy attacks for sensitive XR applications. Furthermore, a key limitation of all these previous works is that they applied DP uniformly to all features without considering the relative importance of each feature to the private DL model's predictions. Such a *blind application* of DP may introduce unnecessary noise to less relevant features, significantly increase inference time, and reduce model accuracy, making it very challenging for real-time deployment in XR, thus hampering the user's immersive experience in XR environments.

*Contribution.* This paper introduces a novel framework combining explainable AI (XAI) and DP-enabled privacy-preserving mechanisms to defend against privacy attacks in AI XR models. Specifically, at first, we develop three DL models, i.e., convolutional neural network (CNN), long short-term memory (LSTM), and Transformer, for three XR applications classifying cybersickness [30], emotions [68], and activity classification [28] of users. We then apply the MIA and RDA against these developed AI XR applications and show that RDA and MIA can cause serious privacy concerns. Next, we leverage a post-hoc XAI method to identify the most dominant features in these AI XR models for applying $\varepsilon$-DP to only those identified features during inference time. For applying $\varepsilon$-DP, we consider both the XR datasets and AI models. This contrasts the SOTA $\varepsilon$-DP that is typically applied only to the data samples blindly to all features, making the dataset private while ignoring the DL part [19, 20, 73]. The combination of XAI and $\varepsilon$-DP provides quantifiable resilience against known XR privacy attacks according to an adjustable privacy budget $\varepsilon$, which provides a strong privacy guarantee, enhancing model utility, reducing inference time, and maintaining high classification accuracy. Finally, we deploy the proposed XAI-guided $\varepsilon$-DP-enabled private model on a consumer-grade XR headset, design a user interface (UI), namely *PrivateXR*, which provides a real-time prediction of task (e.g., cybersickness) severity levels while protecting user privacy during XR gameplay and validate the UI through a user study, as shown in Figure 1.

*Key results.* Our experimental results show that the XAI-guided $\varepsilon$-DP method keeps its promise to protect user privacy against strong adversaries while providing better utility. For instance, the XAI-guided $\varepsilon$-DP AI XR Transformer model successfully reduces the re-identification rate by up to 39%, 33%, and 28% and MIA attack success rate by up to 43%, 48%, and 51% for cybersickness, emotion, and activity classification tasks compared to the non-private Transformer model. Regarding effectiveness, using the XAI-guided $\varepsilon$-DP method leads to up to $\approx 2\times$ speedup in inference time compared to the traditional $\varepsilon$-DP-based (non-XAI-guided) methods for the same classification tasks while maintaining a high classification accuracy of up to 97%, 94%, and 96% for the same tasks. Furthermore, user study results show that 80% of users reported high enjoyment across all privacy settings during gameplay.

## 2 RELATED WORKS

Privacy has been a matter of concern for XR. XR systems collect various sensor measurements such as eye-tracking [7, 52, 77], head-tracking [30], heart rate [68], and more. These data contain sensitive information related to users' biometrics, behaviors, identities, and real-world surroundings. While such data are essential for utility tasks, such as improving the efficiency of XR rendering, they also present privacy risks, especially when enabling targeted advertising based on unique user characteristics, which may be perceived as intrusive. Recently, researchers have begun exploring domain-specific identification [18, 20, 47, 50, 51, 70, 71], which involves identifying a user performing a specific task (e.g., moving objects to a bin [59], throwing a ball [48]) using models trained on different tasks by the same user within the same domain. For instance, authors in [18, 20] applied a re-identification attack using eye-tracking data to identify users viewing different 360° images [18]. In contrast, authors in [54–56] performed a re-identification attack to identify approximately 50,000 users playing the popular Beat Saber game. While Meng et al. [46] applied a de-anonymization-based privacy attack by exploiting users' movement signatures to identify individuals in VRChat. Furthermore, keystroke inference attacks [75] have been demonstrated in shared virtual environments, where an attacker could recover typed content by observing another user's avatar, leading to personal privacy breaches.However, a key limitation of prior works is their focus on raw sensor data privacy while ignoring vulnerabilities in the underlying ML/DL models that use this data for different XR applications. A wide range of ML and DL-based methods have explored eye region detection, gaze estimation, and cognitive state inference for real-time user interaction and rendering [12, 25], with several works also examining the link between gaze behavior and human emotion in virtual environments [26, 43]. For instance, authors in [26] used an ML-based LS-SVM method to classify emotion classes (e.g., arousal, valence, etc.). Beyond emotion, eye movements serve as key indicators of human perception in XR. However, XR HMDs can trigger cybersickness, a perceptual discomfort, motivating DL-based approaches for its detection and mitigation using a variety of XR user data (eye-tracking, head-tracking, etc.) [30–32, 36, 37, 63]. For instance, Jeong et al. [31] employed spatial and temporal Transformer encoders to predict cybersickness from multimodal time-series XR sensor data. In parallel, researchers have explored task recognition using eye movement data [1, 17, 19, 72, 78]. For instance, EHTask [1] employed a DL-based approach to classify tasks (e.g., free viewing, visual search, etc.) from 360-degree VR videos, while Wang et al. [72] proposed the TRCLP model for eye-based task recognition.

Formal privacy methods, particularly DP [11, 19, 20, 34, 35, 38, 41, 56], are widely used in XR to protect sensitive data (e.g., eye tracking) by restricting output distributions. However, these methods often depend on the target XR application and apply privacy mechanisms at the eye-tracking sample levels to make the dataset private. Moreover, it should be noted that eye-tracking data are already noisy sources of information; thus, adding additional noise to preserve privacy may ruin the user experience. In addition, a few works applied reinforcement learning agents [23] for tasks like expertise prediction and document-type classification, model-fitting approaches such as GazeDirector [74] for gaze estimation and redirection tasks without requiring person-specific training data, and federated learning-based methods [10, 22] to preserve data privacy for gaze estimation tasks using pseudo-gradient optimization. However, these works do not provide a formal and provable (cryptographic) privacy guarantee (e.g., DP and secure aggregation). Also, they may not generalize well to broader XR tasks (e.g., user activity identification). Moreover, applying them to high-dimensional XR data poses challenges due to data heterogeneity and trade-offs in maintaining formal guarantees. Recently, Kundu et al. [35] applied DP to DL-based cybersickness classification models under MIA attacks. However, their methods are often application-specific (e.g., cybersickness) and lack generalizability across XR tasks. Another key limitation of their work is the uniform application of DP across all input features, ignoring their relative importance. This degrades model accuracy and increases inference latency, making such methods unsuitable for real-time XR due to reduced responsiveness and user immersion. Meanwhile, XAI methods have shown promise in XR tasks like cybersickness and attention analysis by identifying key features [36, 64]. We argue that leveraging feature importance to selectively apply DP to only the most sensitive features reduces noise, enhances efficiency, and improves practicality for real-time XR deployment.

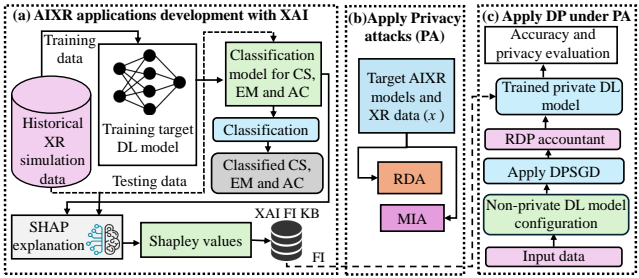

Figure 2: An overview of our proposed XAI-guided DP-enabled privacy protection mechanisms in AI XR applications, designed to protect XR user privacy, where FI represents feature importance scores and KB refers to the knowledge base.

## 3 PROPOSED METHODOLOGY

Our Proposed method is comprised of three phases: (1) Developing DL models for specific applications in XR and applying the XAI method in these AIXR models, (2) Evaluating the privacy leakage of developed DL models for XR by applying MIA and RDA, and (3) Integrating the proposed XAI-guided $\varepsilon$-DP-based mechanisms in AI XR applications to protect the privacy of XR users, as illustrated in Figure 2. The details are as follows.

### 3.1 Developing AI applications in XR

We begin by developing DL models tailored for specific XR applications, focusing on classifying cybersickness, user emotions, and activity. We train three models for this: a CNN, LSTM, and Transformer. We choose these models since they are popular and commonly used in the SOTA XR applications [31, 54, 55, 60]. The LSTM model consists of six 128-unit LSTM layers, a dropout layer with 15% recurrent dropout to mitigate overfitting, and two dense layers with rectified linear unit (ReLU) activation function. The CNN model is composed of two convolutional layers followed by two fully connected layers, each with a filter size of 64, kernel size of 10, and a stride of 1 with input padding of (5,5), a max pooling layer with pool size 2, ReLU activation, a 15% dropout layer, and two dense layers. A Transformer is a sequence-to-sequence model that uses an encoder to map input sequences to high-dimensional representations, which the decoder then transforms into output sequences [69]. In our work, the Transformer network consists of three main components: (1) a positional encoding layer to embed input sequences and encode temporal information; (2) an encoder with multi-head self-attention to capture key patterns in time-series data; and (3) the output layer that predicts cybersickness/emotion/activity through an activation function. The model uses a 64-dimensional attention vector, followed by normalization and a feed-forward MLP with 256 hidden units. We use categorical cross-entropy in our DL-based all classification models as the loss function.

### 3.2 Evaluating the privacy leakage of AI XR applications

In this section, we present the details of our threat model. We focus on two dimensions of privacy leakage in AI XR applications: RDA and MIA. We apply these attacks against these developed XR DL models. We provide the brief information about the threat model for RDA and MIA against XR DL models as follows.

**Membership inference attack:** The main intuition behind MIAs is that the DL models often behave very differently on the training data than test data (e.g., data never seen before). Thus, an adversary can train an DL model denoted as *attack model* to recognize such disparities in the classification model's behavior and then use that model to determine if a XR user data sample was used to train the classification model.

**Re-identification attack:** A RDA occurs when an adversary successfully matches anonymized data to a specific individual by leveraging auxiliary information. Even if personal identifiers (i.e., names and birthdates) are removed from a dataset, remaining attributes such as age, gender, and behavioral data (e.g., eye-tracking) can still enable identification. For instance, if a dataset retains eye-tracking data and demographics, an adversary can select the identities that match the demographics of their target and then train and apply a model to re-identify individuals.

*Threat Model:* The threat models we consider for RDA and MIA in this paper are inspired by [18, 19] and [35], respectively. For the RDA, we assume that an attacker has access to a public XR dataset. The attacker trains an identification model to take XR dataset features (e.g., saccade, fixation, etc.) as input and output the associated identity. Given a new sample from the dataset, the attacker applies the trained model to infer the XR user identity. The RDA attack is successful if the model correctly identifies the individual. To execute the RDA on these XR applications, we make a similar assumption proposed by David et.al. [18, 19]. In contrast, for MIA, we assume that an attacker's primary goal is to determine whether a specific individual's data was part of the training dataset used to train a DL model for prediction tasks in XR to leak XR user privacy. To execute the MIA on DL models in XR, we make a similar assumption proposed by Kundu et.al. [35]. Similar to [35], we assume that the adversary has black-box access to the target model (e.g., emotion recognition), i.e., the adversary can submit a data point to the target model and then obtain the probabilistic output of the target class. Thus, the adversary trains $N$ number of shadow models to achieve a higher attack success rate (ASR) and creates an attack model to query the target model and infer membership. *The main difference between a RDA and an MIA is that the former involves identifying an individual by linking anonymized data to their real-world identity using auxiliary information, such as behavioral patterns or demographic data. In contrast, an MIA determines whether a specific individual's data was included in the training set of a DL model used in XR applications.*

### 3.3 Explainable AI-Guided Feature Selection for DP

Explainable AI (XAI) techniques have been widely adopted to explain various aspects of AI-powered XR applications (e.g., cybersickness, attention, etc.,) [36, 64]. However, limited attention has been given to leveraging XAI to understand privacy implications in AI XR applications, particularly in applying DP in a targeted manner to enhance both user privacy and model accuracy. In this work, we employ the SHapley Additive exPlanations (SHAP)-based post-hoc explanation method to explain the outcomes of the non-private DL models (without DP) for different classification tasks (e.g., cybersickness, emotion, and activity). Specifically, we use the *Deep SHAP* method for computing SHAP values from the non-private DL models. SHAP is a feature importance method that assigns a feature significance value to each prediction, highlighting which features most strongly influence the model's prediction by identifying those that contribute most to the forward pass. SHAP is grounded in Shapley values from cooperative game theory [42], which have been developed initially to divide credit among players in a team fairly. Each input feature is treated like a *"player"*, and SHAP values represent each feature's average marginal contribution across all possible feature combinations. For a given set of input samples (e.g., eye-tracking, head-tracking data, etc.,) and non-private DL models (e.g., LSTM, CNN, and Transformer), SHAP explains predictions by quantifying each feature's contribution. SHAP values indicate how individual feature values influence a model's prediction, increasing or decreasing its output. For instance, given a non-private DL model, the SHAP explanation allows us to know to what extent a feature drove a given prediction (i.e., cybersickness). SHAP supports both global and local interpretations: global explanations rank overall feature importance, while local explanations assess feature impact on specific predictions. Our approach focuses on global explanations to systematically identify non-private DL models' top most influ-ential features. Using SHAP, we extract a ranked list of important features, allowing us to selectively apply $\varepsilon$-DP only to the most critical features rather than applying it to all features indiscriminately during model prediction/inference time. For instance, in our case, we apply $\varepsilon$-DP only to the top 1/4 of ranked features, as identified by SHAP, rather than introducing noise across the entire feature set. This selective DP application significantly improves classification accuracy in private DL models while reducing computational overhead. As a result, our selective DP approach enables faster inference, facilitating practical, real-time deployment on XR headsets for privacy-preserving tasks. By integrating XAI-guided feature selection with $\varepsilon$-DP, we simultaneously enhance user privacy and improve the efficiency of private DL models in XR environments.

### 3.4 Integrating DP to AI XR Models

We employ DP [21] to defend against RDA and MIAs by applying $\varepsilon$-DP to the DL models. First proposed by Dwork et.al. [21], DP is a theoretical definition of privacy that offers a stronger notion of privacy guarantee for output data distribution and computations involving aggregate datasets. According to this principle, one shouldn't be able to tell whether a particular record in an input dataset exists by viewing its output, even in the worst-case scenario where all other entries from the original dataset have been leaked. An assumption on adversary knowledge is unnecessary as the DP guarantee applies to any two datasets that differ by at most one element. The key idea behind DP is that it prevents the model from relying too heavily on any single data point. Instead, the model learns general patterns across the entire dataset. This helps ensure that the inclusion or exclusion of one person's data does not significantly influence the model's behavior or outputs, thereby protecting individual privacy. One of the most common ways to implement DP is by adding a small amount of random noise during the learning process. This noise obscures individual contributions, providing a stronger privacy guarantee and preserving the overall utility of the model. The amount of information that could be revealed is controlled by a parameter known as the privacy budget ($\varepsilon$). A lower $\varepsilon$ indicates stronger privacy protection. Techniques such as adding Laplace or Gaussian noise are commonly used to achieve this, as discussed in [21]. Furthermore, applying the DP mechanism to protect privacy when training DL models on aggregate datasets is also possible. Differentially Private Stochastic Gradient Descent (DPSGD) is a well-known algorithm that uses this idea of DP to provide a privacy guarantee for DL. Thus, to protect AI XR DL models against RDA and MIAs, our differentially private DL model has two major components: (1) DPSGD and (2) moments accountant [35]. The details are as follows.

*Differentially Private Stochastic Gradient Descent (DPSGD):* DPSGD incorporates Gaussian noise with gradient updates in Stochastic Gradient Descent (SGD) to ensure DP during DL model training [6]. Unlike traditional SGD, where gradients directly update parameters, DPSGD first clips gradients to keep their L2-norm within a threshold, preventing any single data point from dominating the model and reducing overfitting. Gaussian noise is added to these clipped gradients to ensure privacy while maintaining utility. During the training, the privacy parameters (e.g., epsilon ($\varepsilon$)) are carefully configured to provide robust privacy guarantees. Note that we use Gaussian noise instead of Laplacian noise because Laplacian noise tends to introduce excessive noise and compromise the utility of the AI model's output [53, 57]. In contrast, Gaussian noise provides better utility while offering meaningful privacy protection, is crucial for usability in XR settings, and is computationally efficient for real-time XR applications [35, 53, 57].

*Moments Accountant:* In DPSGD, calculating the overall privacy cost during training is crucial. Unlike traditional SGD, DPSGD requires multiple iterations of training data, each potentially leaking information and reducing the privacy budget $\varepsilon$. To manage this, DPSGD uses privacy accounting techniques, such as the *moments accountant* [6], which provides tighter privacy bounds by selecting appropriate noise scales and gradient clipping thresholds. This approach allows multiple training iterations without exhausting the privacy budget. We use the Renyi Differential Privacy (RDP) privacy accountant method [3, 49] to track the overall privacy budget and provide better utility in classification tasks similar to the work [35].

## 4 DATASETS & EXPERIMENTAL SETUP

This section provides an overview of the evaluation metrics, experimental setup, and dataset we used in our proposed approach.

### 4.1 Dataset

To validate the effectiveness of the proposed method for privacy utilization under RDA and MIA for cybersickness, emotion recognition, and activity classification tasks, we used the three datasets such as Simulation 2021 [2], Emotions Dataset [68], and Activity recognition [28] datasets. It is worth mentioning that these datasets are popular among XR security/privacy researchers and have been widely used in the community [18–20, 22, 23, 27, 35]. The details of these three datasets are given in the following subsections.

*Simulation 2021:* The Simulation 2021 dataset contains users' data from eye tracking, head tracking, and physiological signals collected from 30 participants (15 male, 15 female; mean age 30.04 years). The participants were immersed in five VR simulations: roller coaster, roadside, beach city, sea voyage, and furniture shop. Eye tracking features include convergence distance, gaze direction (x, y, z), pupil diameter, etc., while head tracking data capture Quaternion Rotation of X, Y, Z, and W axis. The dataset contains four cybersickness severity categories: none, low, medium, and high.

*Emotions Dataset:* The XR emotion recognition dataset [4] contains eye tracking and physiological signals (ECG and GSR) from 34 participants (17 male, 17 female; age range: 18–61, mean = 25.0, SD = 7.65) during exposure to 360° video-based virtual environments via XR headsets. The raw eye tracking features include fixation, saccade, micro-saccade, and blink; ECG signal features are inter-beat interval (IBI), beats per minute (BPM), etc., and GSR signal features are skin conductance level ratio of peaks and time, etc.,. The dataset categorizes user emotions into four quadrants of the circumplex model of effects (CMA): high-arousal positive, low-arousal positive, high-arousal negative, and low-arousal negative.

*Activity Recognition Dataset:* The activity recognition dataset, known as EHTask [1], contains eye and head movement data from 30 participants (18 male, 12 female; mean age: 24.5) across four task conditions while viewing 15 360° XR videos. Each video was watched four times under different activities: free viewing, visual search, saliency, and tracking. Eye movement features include fixations and saccades, while head movement features capture horizontal/vertical velocity and acceleration. The dataset has four activity classes: free viewing, visual search, saliency, and tracking.

*Data Preprocessing:* To ensure consistency across different sensor modalities (e.g., eye tracking, head tracking, EEG, GSR) in the Simulation 2021, EHTask, and VREED datasets, we normalized them using the following formula: $X_{\text{normalized}} = \frac{X-\mu}{\sigma}$. Here, $X_{\text{normalized}}$ represents the normalized sample, $\mu$ is the mean, and $\sigma$ is the standard deviation of the sample. Note that all samples in these datasets were time-synchronized, thus preserving the time dimension of the dataset. To validate our approach, we randomly partitioned each dataset into training and testing sets based on the specific task: 70/30 for cybersickness classification, 75/25 for activity recognition, and 60/40 for emotion recognition.

### 4.2 Experimental Setup

We used TensorFlow-2.4 [24] and PyTorch 2.3 [61] for training and evaluating our non-private DL models. We used the IBM-Adversarial Robustness Toolbox [58] to implement MIA attacks. For explaining the non-private DL models, we used the SHAP [42] library. To apply the DP in DL models, we used Tensorflow privacy [3] and Opacus [76], implementing the DPSGD algorithm with the RDP accountant method for privacy evaluation. All the private and non-private models were trained on an Intel Core i9 Processor and 128GB RAM option with NVIDIA GeForce RTX 3090 Ti GPU.

*Hyper-Parameter:* We used the Adam optimizer with 250 epochs and a batch size of 256 for training non-private DL models, applying a learning rate of 0.001 for non-private and private DL models. We implemented early stopping with a patience value of 30 to prevent overfitting. Note that non-private DL models are those without the DP mechanism, while private DL models include the DP mechanism. We also employed 10-fold cross-validation to train and validate the non-private DL models, partitioning the dataset into $k$ groups (i.e., $k = 10$). The same batch size and learning rate were used for training the attack models as in the non-private condition. We use a Radial Basis Function Network (RBFN) to evaluate identification rates under RDA for AI XR applications, similar to the work [18, 19]. The RBFN, with a single hidden layer, generates class probabilities based on input features and identifies the class with the highest probability as the most likely match. Feature vectors from an unknown individual are classified across multiple stimuli, and prediction scores are averaged to determine identity. Identification rates are calculated as the percentage of individuals correctly classified and averaged over 30 runs with different training/testing splits (70/30, 75/25, and 60/40) for cybersickness, emotion, and activity classification tasks. To execute the MIA on DL models in XR, we follow similar assumptions from [35], which require training shadow models using identical methods as the target model. The adversary trains $N$ shadow models ($N = 20$ for cybersickness, $N = 10$ for emotion, and $N = 15$ for activity classification) to achieve a higher attack success rate (ASR) similar to the work [35]. Once shadow models are trained, the adversary uses an attack model to query the target model and infer membership. We evaluate our defense mechanism against RDA and MIA using the $\varepsilon$-DP strategy for cybersickness, emotion, and activity classification tasks. The DL models were trained for $m$ epochs ($m = 200$ for cybersickness, $m = 100$ for emotion, and $m = 130$ for activity classification) at each noise level, followed by calculating the privacy budget and accuracy of the final epoch.

*Evaluation Metrics:* We use the most widely used performance metrics to evaluate the performance of these models with *balanced accuracy* [13], which is better suited for imbalanced data. To evaluate MIA performance (i.e., ASR), we use *Area Under Characteristics (AUC)* scores, which compare the attack's true-positive rate (TPR) and false-positive rate (FPR) across different decision thresholds, similar to [14, 35]. The AUC score reflects the attacker's ability to distinguish between members and non-members of the target model's training dataset. A higher AUC indicates a more successful attack, whereas a score of 0.5 suggests random guessing and a score below 0.5 implies an unsuccessful attack. Moreover, we use the *mean accuracy* to measure privacy and compare the privacy vs. utility tradeoff, defined as the accuracy when DP is present under the MIA and RDA conditions.

## 5 EXPERIMENTAL RESULTS

This section presents the results of our proposed framework, evaluating the impact of RDA and MIA on AI XR applications and the effectiveness of our proposed XAI-guided $\varepsilon$-DP-based defense against RDA and MIAs.

### 5.1 AI XR/Target Model Accuracy for Cybersickness, Emotion, and Activity Classification

For the cybersickness, emotion, and activity classification task, we train three DL models, namely LSTM, CNN, and Transformer, to classify the level of cybersickness, emotion, and activity in a non-private manner (without DP) and without RDA and MIAs. The cybersickness severity classification using LSTM, CNN, and Trans-former models results in balanced accuracy of 0.94, 0.90, and 0.96, respectively. The emotion classification task using LSTM, CNN, and Transformer models has a balanced accuracy of 0.92, 0.89, and 0.94. On the other hand, the activity classification task using LSTM, CNN, and Transformer models results in a balanced accuracy of 0.94, 0.93, and 0.98. We observe that the overall performance of the Transformer model is better than that of the other models for cybersickness severity, emotion, and activity classification tasks.

Table 1: ASR of identifying the members in the training data regarding AUC score for MIA and re-identification rate for RDA in cybersickness (CS), emotion (EM), and activity (AC) classification tasks.

| Models | MIA AUC score(%) | | | RDA rate (%) | | |
|---|---|---|---|---|---|---|
| | CS | EM | AC | CS | EM | AC |
| LSTM | 81 | 77 | 79 | 40 | 38 | 33 |
| CNN | 74 | 71 | 74 | 44 | 40 | 33 |
| Transformer | 85 | 80 | 83 | 48 | 43 | 36 |

### 5.2 Attack Evaluation: Impact of RDA and MIA on AI XR Models

First, we evaluated identification rates for AI XR applications for cybersickness, emotion recognition, and activity classification tasks using an RBFN similar to the work [18, 19], as discussed in Sections 3.2 and 4.2. Note that identification rates are calculated only from the randomly selected testing data. Table 1 summarizes the re-identification rates of Transformer, LSTM, and CNN models for cybersickness, emotion recognition, and activity classification tasks. We observe that the trained Transformer model achieves the highest re-identification rate for cybersickness, emotion recognition, and activity classification tasks. For instance, in cybersickness classification tasks, the Transformer model achieves a re-identification rate of 48%, which is 1.11×, and 1.35× higher than LSTM and CNN models, respectively. A similar phenomenon is also observed for emotion recognition and activity classification tasks. Furthermore, Table 1 summarizes the success rate of MIA represented using the AUC score under the MIA for the cybersickness, emotion, and activity classification tasks using three DL models (e.g., LSTM, CNN, and Transformer). In our work, ASR measures how accurately an attacker can determine whether XR user data was used to train the AI XR model. Higher ASR values indicate tremendous success for the attacker, meaning they can more accurately identify whether particular data was a member of a cybersickness, emotion recognition, and activity classification training dataset. We observe that the Transformer model has a higher AUC score of 85%, which is 1.05× and 1.16× higher than the LSTM and CNN models. This AUC score indicates that using a Transformer model, the attacker has 85% success in correctly identifying whether XR user data was part of the training set for the Simulations 2021 dataset. We find that the attack's average success rate correlates directly with the target model's perfect testing accuracy, meaning that better models are more vulnerable to MIA attacks. For instance, the Transformer model has higher accuracy in cybersickness classification; thus, this model achieves a higher ASR. However, that is not true for other models, such as CNN, which has limited classification results (low accuracy). A similar phenomenon is also observed for emotion and activity classification tasks.

### 5.3 Explainable AI-Guided Feature Selection for DP in AI XR Applications

Figure 3 shows the overall feature importance for cybersickness, emotion, and activity classification tasks using the Transformer model. We apply the SHAP-based global explanations method to explain the Transformer model prediction outcome. A shapely value is used to calculate the ranking of the most important features contributing to the different XR classification tasks, with important features at the top and the least important ones at the bottom. A large absolute Shapley value indicates that the feature has a higher average impact on model output than those with smaller absolute values. For instance, in cybersickness classification tasks, we observe that features such as *Normal Right Eye Origin X, Y, and Z axis*, Normal Left Eye Origin Y axis,*Gaze Origin World Spacc Y axis* etc., are the most influential features in cybersickness severity classification for the Simulation2021 dataset, as shown in Figure 3 (a). It is observed that eye-tracking features have a much more significant impact on the cybersickness classification than physiological and head-tracking features. The reason is that eye tracking features contain insightful information such as the type of blink of the user, gaze behavior, and the position of the pupil to track the user's activity [36]. Therefore, these eye-tracking features can influence the level of cybersickness, thus enhancing the model prediction. Similarly, for the emotion recognition task, it is observed that features such as *microsaccades peak velocity,microsaccades direction, saccades direction, blink duration*, etc., have a much stronger influence in the emotion recognition for the emotion dataset, as shown in Figure 3 (b). We observe that for the emotion tasks, most of the features contributing to model prediction are, again, eye-tracking features. A similar phenomenon is also observed for the activity classification task (i.e., eye-tracking features are the most influential features) and other DL models. Importantly, we observe that SHAP value rankings remain stable across different tasks. For instance, eye-tracking features consistently emerged as the most influential across all classification tasks. This is likely because eye-tracking data captures rich information inherently correlated with XR user states and behaviors. Such stability in feature importance reinforces the robustness of our XAI-guided approach by ensuring that the same key features are consistently identified as the most critical for model predictions. Thus, identifying these features via SHAP explanation collectively offers a detailed picture of which features truly impact model predictions, allowing us to apply DP only to those top features, preserving model privacy while maintaining utility.

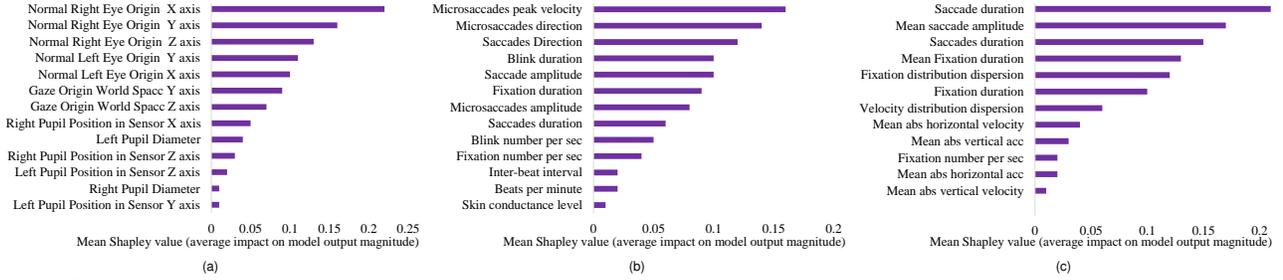

Figure 3: Overall feature importance using SHAP for the Transformer model in (**a**) cybersickness, (**b**) emotion, and (**c**) activity classification tasks, highlighting the most influential features for each task.

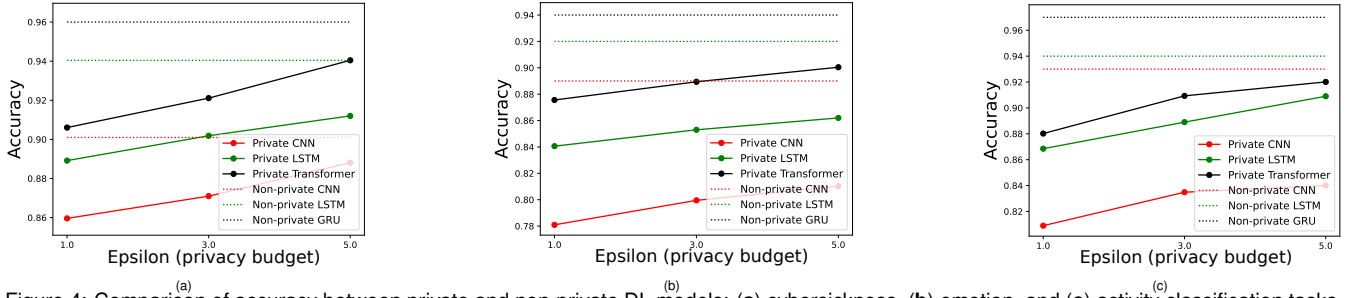

Figure 4: Comparison of accuracy between private and non-private DL models: (**a**) cybersickness, (**b**) emotion, and (**c**) activity classification tasks.

### 5.4 DP for Defense Against MIA and RDA in AI XR applications

This section evaluates our without XAI and with XAI-guided defense mechanism against RDA and MIA using DP for AI XR applications through their *privacy-utility-trade-off* evaluation.

***Privacy-Utility-Trade-off Evaluation for XR classification tasks:*** Figure 4 shows the mean accuracy of cybersickness severity, emotion, and activity classification using three DL methods (e.g., LSTM, CNN, and Transformer) for different privacy budgets $\varepsilon$ and baseline conditions (without privacy budgets, $\varepsilon$ which are denoted as non-private models) under MIA and RDA using traditional $\varepsilon$-DP-based (non-XAI-guided) feature selection method. In this setup, DP is uniformly applied to all input features. For instance, in the cybersickness classification task, DP is applied equally across eye-tracking, head-tracking, and physiological features data. For this, we use three privacy budget groups (similar to the privacy budget groups in [35, 56]), i.e., high ($\varepsilon = 1$), medium ($\varepsilon = 3$), and low ($\varepsilon = 5$). We use the privacy parameters $\delta = 0.00007, 0.00005, and 0.00008$ for the cybersickness, emotion, and activity classification tasks. We use the same rule of thumb to measure the $\delta$, which is set to be less than the inverse of the training data size, similar to the work [6]. We observe that private DL models, specifically the private Transformer model, achieve higher accuracy in private and non-private settings than other DL models for these classification tasks. We observe that relaxing the $\varepsilon$ and parameters $\delta$ significantly increases the classification task performance of the private models. For instance, in the cybersickness classification task, the Transformer model achieves a mean accuracy of 0.94 with a $\varepsilon = 5$ and $\delta = 0.00007$, as shown in Figure 4(a). In contrast, it is observed that there is a considerable accuracy drop for other DL models with low $\varepsilon$. A similar phenomenon is also observed for emotion and activity classification tasks.

***MIA and RDA attacks evaluation under DP without XAI:*** We also evaluate how much the private models leak information about their training data under MIA and how private models correctly identify individuals who were correctly classified with the largest class score under an RDA using traditional $\varepsilon$-DP-based (non-XAI-guided) feature selection method. This evaluation provides a baseline for understanding the limitations of applying DP without guidance from the SHAP-based explainability method, which provides each feature's relative importance to the DL model's predictions. Specifically, we apply $\varepsilon$-DP uniformly to all input features, regardless of their individual contribution to the model's predictions. Table 2 shows the identification rate of RDA and ASR of MIA in the non-private and private classification models (e.g., Transformer, LSTM, and CNN), along with the private classification mean accuracy (MA) for the private models for different $\varepsilon$. To do this, we use the highest $\varepsilon$ value (i.e., $\varepsilon$=5) for different DL models, which provide the highest mean accuracy. In our work, we first evaluate our proposed method with private data (i.e., adding $\varepsilon$-DP only in the sample levels) and a non-private model (i.e., without $\varepsilon$-DP) for both attacks. Then, we evaluate our method with non-private data (without $\varepsilon$-DP) with the private models, and finally, we evaluate our method with both $\varepsilon$-DP-enabled private data and private models. We observe that the re-identification rate of RDA and ASR of MIAs against the corresponding $\varepsilon$-DP-enabled target private models is not significant and is lower than the baseline non-private models and private data with non-private models. For instance, for the Transformer model in the cybersickness classification task, we observe that the re-identification rate and ASR against the corresponding $\varepsilon$-DP-enabled target private models are insignificant and lower than the baseline non-private models. For instance, the private Transformer model with private data achieves a re-identification rate of 14%, 16%, and 15% for cybersickness, emotion, and activity classification tasks, which reduces the identification rate up to 39%, 33%, and 28% than the non-private Transformer model. While, for the MIA attack in the cybersickness classification task, the private Transformer model achieves an AUC score of 43%, 48%, and 51% for cybersickness, emotion, and activity classification tasks, which reduces the ASR up to 39%, 31%, and 44% than the private data with the non-private Transformer model, 21%, 17%, and 23% than the non-private data with the private Transformer model and 50%, 45%, and 51% than the non-private Transformer model. These low AUC scores (less than 50%) for MIA reflect stronger privacy protection, as the model does not leak distinguishable information about its training data. Furthermore, from the utility perspective, our results show that the proposed defense has minimal effect on utility. For instance, the private AI XR Transformer model achieves a classification accuracy of up to 94%, 90%, and 92% for the same classification tasks. We also observe that the theoretical $\varepsilon$ privacy guarantee seems to have a strong correlation with the MIA success rate and identification rate for RDA, where, most of the time, a value of $\varepsilon$ is associated with a more significant decrease in ASR for MIA and identification rate for RDA. This result, therefore, serves as evidence that the DP method promises to offer protection against sophisticated attack methods such as the MIA and RDA, thus protecting the privacy of the XR simulation data and DL-based models in classification tasks when offering an acceptable utility level. A similar phenomenon is also observed for emotion and activity classification tasks.

*MIA and RDA attacks evaluation under DP with XAI:* We also evaluate how much the private models leak information about their training data under MIA and how private models correctly identify individuals who were correctly classified with the largest class score under an RDA using XAI-guided $\varepsilon$-DP-based feature selection method. In this setup, we first use SHAP-based identified most influential features contributing to model predictions across classification tasks, as discussed in Section 5.3. $\varepsilon$-DP is then selectively applied only to these top-ranked features rather than uniformly across all features of the input data. This targeted approach minimizes the amount of $\varepsilon$-DP injected into the model, providing better utility (higher classification accuracy) and faster inference times while having minimal/no impact on ASR for RDA and MIA. Note our XAI-guided $\varepsilon$-DP evaluation method did not consider other conditions, such as no privacy and *NPD + PM*), as our method focuses solely on applying $\varepsilon$-DP during the prediction/inference phase using private data. Thus, we only consider two conditions (e.g., *PD + NPM* and *PD +PM*). Table 2 shows the identification rate of RDA and ASR of MIA in the private data with non-private model and private data with private model. We observe that the re-identification rate of RDA and ASR of MIA against the corresponding $\varepsilon$-DP-enabled target private models is insignificant compared to those traditional $\varepsilon$-DP-based (non-XAI-guided) method. However, the model utility and inference time significantly increase compared to the traditional $\varepsilon$-DP-based (non-XAI-guided) method. For instance, the XAI-guided $\varepsilon$-DP-enabled private CNN model achieves a mean accuracy of 92%, 90%, and 93% for cybersickness, emotion, and activity classification tasks, which improves the mean accuracy 4.5%,11.2%, and 10.8% for the same classification tasks than traditional $\varepsilon$-DP-based (non-XAI-guided) methods. A similar phenomenon is observed in other private models and classification tasks.

*Effectiveness:* Furthermore, inference time for private classification tasks without XAI and with XAI-guided $\varepsilon$-DP method is shown in Table 4. It is observed that private classification tasks inference time is significantly lower for the XAI-guided $\varepsilon$-DP method than traditional $\varepsilon$-DP-based (non-XAI-guided) method. For instance, using the XAI-guided $\varepsilon$-DP-enabled Transformer model, the inference speed per sample takes 1.35 ms, 1.05 ms, and 1.45 ms on the CPU for cybersickness, emotion, and activity classification tasks **leading to an up to $\approx 2\times$ speed up in inference than traditional $\varepsilon$-DP-based (non-XAI-guided) method for the same tasks**. A similar phenomenon is observed for other private DL models.

## 6 PRIVATEXR UI DEVELOPMENT

This section explains the details of implementing our proposed PrivateXR (XAI-guided $\varepsilon$-DP-enabled private model) UI development for a specific use case, i.e., the AI-based cybersickness prediction in the XR roller coaster simulation. This UI allows XR users to engage and interact with the private cybersickness prediction model for different privacy levels (e.g., low, medium, and high), offering real-time predictions of cybersickness severity (e.g., none, low, medium, and high) while protecting user privacy during gameplay. Developed as a Unity (*C#*) plugin, PrivateXR can seamlessly integrate into any XR application using MelonLoader [45]. It is worth mentioning that while our current focus is on cybersickness prediction, the method is generalizable and can be extended to other privacy-sensitive classification tasks in XR (e.g., emotion and activity classification).

*Virtual Environment and Apparatus:* The virtual environment was an XR roller coaster that allowed different locomotion, 3D object manipulation, and visuals. For developed PrivateXR, we used an HTC-Vive Pro Eye headset to render the XR roller coaster simulation, which offers a display resolution of $2880 \times 1600$ pixels per eye and a refresh rate of 90Hz. To capture eye-tracking and head-tracking data, we utilize the HTC SRanipal SDK and Tobii HTC Vive Devkit API [5]. The XR roller coaster simulation was rendered using a computer equipped with an Intel Core i9 CPU with 128 GB of memory and an Nvidia GeForce RTX 3090 GPU.

*Private DL model deployment and Inference phase in the UI:* After training and validation, the non-private (without DP) and private DL model (with XAI-guided $\varepsilon$-DP) is converted to ONNX format [40] and deployed on an HTC Vive headset. In the inference phase, the pre-trained deployed non-private and private model predicts the severity level of the cybersickness (e.g., none, low, medium, and high) using streaming XR simulation XAI-guided $\varepsilon$-DP-enabled private eye and head tracking data for the private model and without $\varepsilon$-DP-enabled eye and head tracking data for the non-private model.

### 6.1 Settings of User Interface

The main objective of the PrivateXR is to provide a real-time prediction of cybersickness severity levels (e.g., none, low, medium, and high) while protecting user privacy during XR gameplay. More specifically, it predicts cybersickness severity levels simultaneously in real-time for both non-private and private settings and visualizes them as a bar with different colors (red means high, yellow means medium, green means low, and gray means none severity levels) to indicate user different severity levels. The flexible interface is shown in Figure 1 (a), which has two modes: no privacy (without DP) and privacy (with DP). The PrivateXR interface is shown in Figure 5, where XR users select low privacy level mode from the privacy switch, which provides a real-time prediction of cybersickness severity level (e.g., high cybersickness (CS) severity) while providing lower privacy but highest utility during gameplay (XR roller coaster). Specifically, we expose the following options:

*Prediction Toggle:* The *No privacy* switch provides real-time visual feedback of the predicted internal state of the user's cybersickness severity levels via a non-private deployed DL model without DP. The *Privacy* switch provides real-time visual feedback of the predicted internal state of the user's cybersickness severity levels via an XAI-guided $\varepsilon$-DP-enabled private deployed DL model for three privacy settings: 1) *Low* privacy provides lower privacy while providing better utility (high accuracy for cybersickness classification tasks), 2) *Medium* privacy provide moderate privacy and moderate utility, and 3) *High* privacy provide higher privacy but lower utility.

Table 2: Re-identification attack (RDA) identification rate and MIA attack success rate (ASR) in terms of the AUC score of the non-private classification (CL) models denoted as *No privacy*, RDA and MIA ASR along with the DP-based private data (PD) for non-private model (NPM) and PD along with private models (PM) classification mean accuracy (MA) for cybersickness, emotion, and activity classification tasks. The notation X / Y / Z represents the Transformer model/ LSTM model / CNN model.

| CL tasks | No privacy(%) | | | PD + NPM (%) | | | No PD + PM (%) | | | PD + PM (%) | | |
|---|---|---|---|---|---|---|---|---|---|---|---|---|
| | MIA | RDA | MA | MIA | RDA | MA | MIA | RDA | MA | MIA | RDA | MA |
| CS | 85/81/74 | 48/40/44 | 96/94/90 | 71/70/67 | 14/16/15 | 94/92/90 | 54/56/59 | 48/40/44 | 94/92/89 | 43/48/51 | 14/16/15 | 94/92/89 |
| EM | 80/77/71 | 43/40/38 | 94/94/89 | 63/60/58 | 15/17/19 | 91/89/84 | 53/54/56 | 43/40/38 | 90/88/82 | 44/48/50 | 15/17/19 | 90/86/81 |
| AC | 83/79/74 | 36/30/27 | 98/94/93 | 72/65/62 | 11/14/19 | 93/91/88 | 52/55/58 | 36/30/27 | 93/90/86 | 40/46/48 | 11/14/19 | 92/90/84 |

Table 3: RDA identification rate and MIA ASR in terms of the AUC score of the DP-based PD for NPM and PD along with PM classification mean accuracy (MA) for cybersickness, emotion, and activity classification tasks with XAI guided. The notation X / Y / Z represents the Transformer model/ LSTM model / CNN model.

| CL tasks | PD + NPM (%) | | | PD + PM (%) | | |
|---|---|---|---|---|---|---|
| | MIA | RDA | MA | MIA | RDA | MA |
| CS | 71/73/68 | 14/16/15 | 98/95/93 | 43/49/52 | 14/16/15 | 97/94/92 |
| EM | 65/60/59 | 15/17/19 | 96/93/91 | 42/48/52 | 15/17/19 | 94/93/90 |
| AC | 73/67/65 | 11/14/19 | 97/94/94 | 40/47/50 | 11/14/19 | 96/94/93 |

Table 4: Inference time in mili seconds (ms) for per sample comparison of private DL models for classification tasks with and without XAI-guided features selection for applying DP. The notation X/Y represents the feature selection without XAI/with XAI for applying DP.

| CL tasks | CNN | LSTM | Transformer |
|---|---|---|---|
| CS | 2.31/1.32 | 2.36/1.35 | 2.38/1.35 |
| EM | 2.04/1.02 | 2.08/1.03 | 2.13/1.05 |
| AC | 2.44/1.4 | 2.51/1.44 | 2.57/1.45 |

*Using the UI:* Here, we present how users choose each of the Toggles: 1) Access the Settings: Navigate to the settings menu from the main interface, 2) No privacy Toggle use for enabling or disabling no privacy mode and privacy Toggle use for enabling or disabling different privacy mode (e.g., low, medium, and high).

*UI Evaluation:* This section presents the detailed results of our proposed and developed UI (*PrivateXR*) evaluation. Specifically, we explain the deployed private and non-private DL models in the developed UI in detail. We used the trained Transformer-based private and non-private DL models (since Transformer performs the best in terms of both non-private and private prediction tasks as shown in Figure 4 and Tables 2 and 3). We deployed both models in the HTC Vive Pro headset to validate the UI. The total size of the deployed Transformer models is 4.89 MB for the non-private version and 4.99 MB for each private model. For the private Transformer models, we trained three separate models corresponding to different privacy budgets (low, medium, and high) to offer varying levels of privacy protection. We integrated a privacy level slider within the UI, allowing users to control the trade-off between privacy and classification accuracy by dynamically adjusting the DP parameter ($\varepsilon$) in real time. When a user selects a low privacy setting, the model trained with a low privacy budget is used for prediction. The training time for the deployed non-private and private Transformer models was 422 and 489 minutes, respectively. In the XR simulation, the deployed non-private Transformer model requires only 0.00018 seconds, while the private Transformer model (without XAI-guided $\varepsilon$-DP) takes 0.0040 seconds, and the XAI-guided $\varepsilon$-DP-enabled private Transformer model takes 0.00024 seconds to predict cybersickness severity per frame. The XAI-guided $\varepsilon$-DP method, which applies $\varepsilon$-DP to selective features during real-time prediction, significantly reduces inference time compared to the non-XAI method while protecting XR user privacy. We only include the traditional $\varepsilon$-DP-based (non-XAI-guided) method to evaluate efficiency. Otherwise, our developed UI does not offer this method as an option in the privacy mode for different privacy levels (e.g., low, medium, and high).

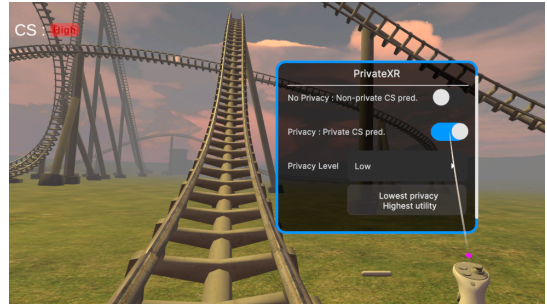

Figure 5: PrivateXR UI in XR roller coaster enables real-time cybersickness prediction with dynamic privacy control at low privacy levels.

## 7 USER STUDY DESIGN TO EVALUATE PRIVATEXR

We conducted a user study to evaluate the impact of the PrivateXR UI, which enables interaction with a private cybersickness prediction model at varying privacy levels, on perceived utility and user experience during XR gameplay. Our goal was to understand how different privacy settings in PrivateXR affect user interactions and overall experience. The study details are provided below.

*Participants, Study Task, and Procedure:* Twelve volunteer participants who received no reward (7 males and 5 females, aged 19 to 39 years, mean age 22.4, standard deviation (SD) 4.23) participated in the study. Our user study was approved by the Institutional Review Board (IRB) at the authors' university. Each participant provided an IRB-approved consent form and demographic information, including age, gender, and prior immersive XR experience. We briefed them on the study procedures and requirements before starting. Participants experienced a seated XR roller coaster simulation with immersive locomotion, 3D object manipulation, and visual elements. It is important to note that the eye and head positions were calibrated before starting the simulation. Upon satisfactory calibration, participants were immersed in an XR trial session. Then, as discussed in section 6, to assess the impact of different privacy settings in PrivateXR on real-time cybersickness predictions, participants experienced four privacy settings, each lasting 120 seconds: 1) No Privacy (NOP), 2) Lowest Privacy-Highest Utility (LPHU), 3) Medium Privacy-Medium Utility (MPMP), and 4) Highest Privacy-Lowest Utility (HPLU). The order of these settings was randomized for each participant. Participants were free to stop at any time for any reason (e.g., if they felt highly uncomfortable) during the experiment. After each setting, we recorded subjective game enjoyment (5-item Likert scale), and qualitative feedback.

*Study Evaluation Results:* To assess how different privacy settings in PrivateXR affect user enjoyment, we asked participants to rate their experience after each session. They responded to the statement: "You enjoyed the XR game in this session," using a 5-point Likert scale, where 5 meant *"Agree"*, and 1 meant *"Disagree"*. As shown in Figure 6, participants reported high enjoyment across all privacy settings during gameplay, with at least 80% giving a rating of 4 or higher. Notably, there is no observable decline in enjoyment when moving from the no privacy setting to the low privacy setting, and only a modest reduction when switching to the high privacy mode. These results show that PrivateXR DP noise does not affect users' utility during gameplay. Furthermore, the qualitative feedback we collected from participants echoed our quantitative findings. Few users reported difficulty distinguishing between the no privacy and low privacy settings. For instance, one participant (**P8**) noted, *"The no privacy and low privacy settings are almost the same for me."* Most participants found that the high privacy settings during gameplay introduced stronger DP noise, which diminished their enjoyment of the game, impacting the overall user experience.

## 8 DISCUSSION

Our experiment's results show that the DP method keeps its promise to protect user privacy against strong adversaries by providing better utility. Specifically, the proposed XAI-guided $\varepsilon$-DP-enabled private AI XR Transformer model reduces re-identification rates by up to 39%, 33%, and 28%, and MIA attack success rates by up to 43%, 48%, and 51% for cybersickness, emotion, and activity classification tasks, respectively, compared to the non-private model. This indicates that attackers are less successful in identifying specific data as members of the training datasets. Additionally, the XAI-guided selective $\varepsilon$-DP method enhances classification accuracy and reduces computational overhead. In terms of utility, the XAI-guided $\varepsilon$-DP-enabled model achieves classification accuracies of 97%, 94%, and 96% for the same tasks. Moreover, the XAI-guided $\varepsilon$-DP method leads to (up to $\approx 2\times$) speedup in inference time compared to the traditional $\varepsilon$-DP-based method for the same tasks. Notably, the re-identification rate and ASR under RDA for MIA remain nearly the same as those under the XAI-guided DP method, showing that it maintains strong privacy protection while significantly improving model utility and reducing inference time, as shown in Tables 2, 3, and 4. Furthermore, our study results showed that users reported high enjoyment across all privacy settings in the PrivateXR UI.

Prior works have focused on preserving the privacy of eye-tracking data in AR/VR/MR devices, offering formal privacy guarantees at the sample level to protect against either RDA or MIA attacks. In contrast, our work identifies the privacy risk in the AI XR models and protects user privacy under RDA and MIA by applying XAI-guided $\varepsilon$-DP while maintaining a high classification accuracy. Moreover, our proposed method provides a privacy guarantee in both sample levels and DL model prediction time under strong adversarial conditions. Previous studies have shown that eye-tracking and head-tracking data are particularly sensitive, potentially revealing private user attributes such as personal identity, gender, and sexual preferences [19, 20, 73]. Our study also found that eye-tracking, head-tracking data, and other physiological data (e.g., HR, EDA, etc.) are *the most sensitive data for leaking XR user privacy in a non-private setting*. However, our XAI-guided $\varepsilon$-DP method significantly reduces the ASR while maintaining strong utility. We observed that private DL models, such as Transformer models, achieve higher accuracy (better utility) for different privacy budgets ($\varepsilon$) in XR classification tasks and outperform several SOTA works in both private and non-private XR cybersickness, emotion, and activity classification tasks [4, 19, 28, 35]. For example, using the Simulations 2021 dataset, Kundu et al. [30] achieved 91% and 89% accuracy with private LSTM and GRU models under high $\varepsilon$. In contrast, our XAI-guided $\varepsilon$-DP-enabled Transformer model attained 98% accuracy with a low $\varepsilon$. Furthermore, on the EHTask dataset for activity classification, David et al. [19] reported 61.8% accuracy using a non-private conditional variational autoencoder with k-same-synth, and 76% (with $\varepsilon = 10$) using the kal$\varepsilon$-ido mechanism, which only provides sample-level privacy for eye-tracking data against RDA, while neglecting the DL model. Similarly, Li et al. [38] applied the kal$\varepsilon$-ido-based privacy mechanism to raw eye gaze data in real-time, focusing solely on sample-level protection, and neglecting potential vulnerabilities (e.g., MIA) in the DL model. In contrast, using the same dataset, our proposed private Transformer model achieved 95% classification accuracy with a low privacy budget ($\varepsilon = 5$) while offering privacy protection for both the sample level and the DL model prediction time. These results demonstrate that our XAI-guided $\varepsilon$-DP method offers strong privacy protection against adversaries while maintaining high utility across diverse XR datasets and tasks. Furthermore, the *PrivateXR* UI can be adapted to other XR applications, such as gaze prediction and stress detection that rely on multimodal XR data (e.g., eye-tracking, head and hand motion) and DL-based methods [12, 19, 54].

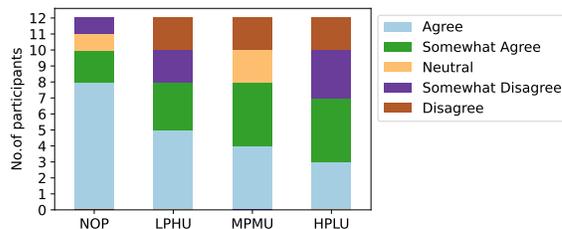

Figure 6: User feedback on game enjoyment across different privacy settings of the PrivateXR UI, rated on a 5-point Likert scale.

## 9 LIMITATION AND FUTURE WORKS

Our proposed framework has a few limitations. Firstly, this work did not incorporate more diverse datasets with additional sensor modalities such as hand tracking [54], head motion [66], and full-body tracking [56], which capture highly sensitive biometric data. For example, hand-tracking can reveal detailed behavioral and emotional cues. Additionally, recent AI XR studies have used advanced models like TimeLLM [16] for cybersickness prediction and GAN [15], MT-CNNs [8] for emotion recognition. Extending our method to such models and tasks could offer deeper insights into the privacy–utility trade-off. Lastly, we could not conduct a large-scale study due to time, resource, and budget constraints. However, in future work, we plan to conduct a more extensive user study with a larger and more demographically diverse participant pool and report the findings with a more comprehensive statistical analysis.

## 10 CONCLUSIONS

This work proposed an XAI-guided $\varepsilon$-DP-enabled framework to protect user privacy against MIA and RDA attacks for AI-based XR emotion recognition, cybersickness, and activity classification tasks. Using three open-source datasets, we demonstrated that such attacks pose significant privacy risks and showcased the effectiveness of our defense approach for these AI XR applications. Specifically, our XAI-guided $\varepsilon$-DP based defense reduced the success rate of MIA and RDA attacks against the Transformer model trained for cybersickness, emotion, and activity classification tasks by 39%, 33%, and 28% for RDA and 43%, 48%, and 51% for MIA. Furthermore, the XAI-guided $\varepsilon$-DP-enabled Transformer model achieved classification accuracies of 97%, 94%, and 96% for the same tasks, exhibiting great utility. Furthermore, we deployed private and non-private DL models on the HTC Vive Pro headset, designed the *PrivateXR* UI, and validated it through a user study. This UI enables XR users to interact with the private cybersickness prediction model and select different privacy levels (low, medium, high), providing real-time predictions of cybersickness severity while ensuring privacy during gameplay. Experimental results show that the XAI-guided $\varepsilon$-DP method significantly reduces inference time (up to $\approx 2\times$) compared to the traditional $\varepsilon$-DP approach while maintaining high utility. Furthermore, 80% of users expressed high levels of enjoyment during gameplay across all privacy settings while using the UI.

## 11 ACKNOWLEDGEMENT

This material is based upon work supported by the National Science Foundation (NSF) under Award Number CNS-2114035 and partially supported by the Army Research Office (ARO) under award number W911NF-23-1-0401. Any opinions, findings, conclusions, or recommendations expressed in this publication are those of the authors and do not necessarily reflect the views of the NSF or ARO.

## REFERENCES

[1] Ehtask dataset. https://zhiminghu.net/hu22_ehtask.html. Accessed: 2024-19-03. 2, 4

[2] Simulation 2021 dataset. https://sites.google.com/view/savelab/research. Accessed: 2024-12-03. 4

[3] Tensorflow privacy. https://github.com/tensorflow/privacy. Accessed: 2023-15-08. 4, 5

[4] Vreed dataset. https://www.kaggle.com/datasets/lumaatabbaa/vr-eyes-emotions-dataset-vreed. Accessed: 2024-19-03. 4, 9

[5] T. AB. Tobii pro sdk. https://www.tobiipro.com/product-listing/tobii-pro-sdk/, 2021. Accessed: March 2, 2023. 7

[6] M. Abadi, A. Chu, I. Goodfellow, H. B. McMahan, I. Mironov, K. Talwar, and L. Zhang. Deep learning with differential privacy. In *Proceedings of the 2016 ACM SIGSAC conference on computer and communications security*, pp. 308–318, 2016. 4, 6

[7] M. Akkoyun, K. Koçoğlu, H. Eraslan Boz, P. Keskinoğlu, and G. Akdal. Saccadic eye movements in patients with mild cognitive impairment: A longitudinal study. *Journal of Motor Behavior*, pp. 1–19, 2023. 2

[8] M. Anand and S. Babu. Multi-class facial emotion expression identification using dl-based feature extraction with classification models. *International Journal of Computational Intelligence Systems*, 17(1):25, 2024. 9

[9] T. Armstrong and B. O. Olatunji. Eye tracking of attention in the affective disorders: A meta-analytic review and synthesis. *Clinical psychology review*, 32(8):704–723, 2012. 1

[10] A. Banse, J. Kreischer, et al. Federated learning with differential privacy. *arXiv preprint arXiv:2402.02230*, 2024. 2

[11] E. Bozkir, O. Günlü, W. Fuhl, R. F. Schaefer, and E. Kasneci. Differential privacy for eye tracking with temporal correlations. *Plos one*, 16(8):e0255979, 2021. 1, 2

[12] E. Bozkir, S. Özdel, M. Wang, B. David-John, H. Gao, K. Butler, E. Jain, and E. Kasneci. Eye-tracked virtual reality: A comprehensive survey on methods and privacy challenges. *arXiv preprint arXiv:2305.14080*, 2023. 1, 2, 9

[13] K. H. Brodersen, C. S. Ong, K. E. Stephan, and J. M. Buhmann. The balanced accuracy and its posterior distribution. In *2010 20th international conference on pattern recognition*, pp. 3121–3124. IEEE, 2010. 5

[14] N. Carlini, S. Chien, M. Nasr, S. Song, A. Terzis, and F. Tramer. Membership inference attacks from first principles. In *2022 IEEE Symposium on Security and Privacy (SP)*, pp. 1897–1914. IEEE, 2022. 5

[15] X. Chen, L. Xu, H. Wei, Z. Shang, T. Zhang, and L. Zhang. Emotion interaction recognition based on deep adversarial network in interactive design for intelligent robot. *IEEE Access*, 7:166860–166868, 2019. 9

[16] Y. Choi, D. Jeong, B. Kim, and K. Han. Early prediction of cybersickness in virtual reality using a large language model for multimodal time series data. In *Companion of the 2024 on ACM International Joint Conference on Pervasive and Ubiquitous Computing*, pp. 25–29, 2024. 9

[17] A. Coutrot, J. H. Hsiao, and A. B. Chan. Scanpath modeling and classification with hidden markov models. *Behavior research methods*, 50(1):362–379, 2018. 2

[18] B. David-John, K. Butler, and E. Jain. For your eyes only: Privacy-preserving eye-tracking datasets. In *2022 Symposium on Eye Tracking Research and Applications*, pp. 1–6, 2022. 1, 2, 3, 4, 5

[19] B. David-John, K. Butler, and E. Jain. Privacy-preserving datasets of eye-tracking samples with applications in xr. *IEEE Transactions on Visualization and Computer Graphics*, 29(5):2774–2784, 2023. 1, 2, 3, 4, 5, 9

[20] B. David-John, D. Hosfelt, K. Butler, and E. Jain. A privacy-preserving approach to streaming eye-tracking data. *IEEE Transactions on Visualization and Computer Graphics*, 27(5):2555–2565, 2021. 1, 2, 4, 9

[21] C. Dwork. Differential privacy. In *International colloquium on automata, languages, and programming*, pp. 1–12. Springer, 2006. 4

[22] M. Elfares, Z. Hu, P. Reisert, A. Bulling, and R. Küsters. Federated learning for appearance-based gaze estimation in the wild. In *Annual Conference on Neural Information Processing Systems*, pp. 20–36. PMLR, 2023. 2, 4

[23] W. Fuhl, E. Bozkir, and E. Kasneci. Reinforcement learning for the privacy preservation and manipulation of eye tracking data. In *Artificial Neural Networks and Machine Learning–ICANN 2021: 30th International Conference on Artificial Neural Networks, Bratislava, Slovakia, September 14–17, 2021, Proceedings, Part IV 30*, pp. 595–607. Springer, 2021. 2, 4

[24] A. Géron. *Hands-on machine learning with Scikit-Learn, Keras, and TensorFlow*. " O'Reilly Media, Inc.", 2022. 4

[25] S. Ghosh, A. Dhall, M. Hayat, J. Knibbe, and Q. Ji. Automatic gaze analysis: A survey of deep learning based approaches. *arXiv preprint arXiv:2108.05479*, 2021. 1, 2

[26] Y.-L. Hsu, J.-S. Wang, W.-C. Chiang, and C.-H. Hung. Automatic ecg-based emotion recognition in music listening. *IEEE Transactions on Affective Computing*, 11(1):85–99, 2017. 2

[27] M. Hu, Z. Luo, Y. Zhou, X. Liu, and D. Wu. Otus: A gaze model-based privacy control framework for eye tracking applications. In *IEEE INFOCOM 2022-IEEE Conference on Computer Communications*, pp. 560–569. IEEE, 2022. 4

[28] Z. Hu, A. Bulling, S. Li, and G. Wang. Ehtask: Recognizing user tasks from eye and head movements in immersive virtual reality. *IEEE Transactions on Visualization and Computer Graphics*, 2021. 2, 4, 9

[29] Z. Hu, S. Li, C. Zhang, K. Yi, G. Wang, and D. Manocha. Dgaze: Cnn-based gaze prediction in dynamic scenes. *IEEE transactions on visualization and computer graphics*, 26(5):1902–1911, 2020. 1

[30] R. Islam, K. Desai, and J. Quarles. Cybersickness prediction from integrated hmd's sensors: A multimodal deep fusion approach using eye-tracking and head-tracking data. In *2021 IEEE International Symposium on Mixed and Augmented Reality (ISMAR)*, pp. 31–40. IEEE, 2021. 2, 9

[31] D. Jeong and K. Han. Precyse: Predicting cybersickness using transformer for multimodal time-series sensor data. *Proceedings of the ACM on Interactive, Mobile, Wearable and Ubiquitous Technologies*, 8(2):1–24, 2024. 2, 3

[32] D. Jeong, S. Paik, Y. Noh, and K. Han. Mac: multimodal, attention-based cybersickness prediction modeling in virtual reality. *Virtual Reality*, pp. 1–16, 2023. 2

[33] B. John, S. Koppal, and E. Jain. Eyeveil: degrading iris authentication in eye tracking headsets. In *Proceedings of the 11th ACM Symposium on Eye Tracking Research & Applications*, pp. 1–5, 2019. 1

[34] B. John, A. Liu, L. Xia, S. Koppal, and E. Jain. Let it snow: Adding pixel noise to protect the user's identity. In *ACM Symposium on Eye Tracking Research and Applications*, pp. 1–3, 2020. 2

[35] R. K. Kundu and K. A. Hoque. Preserving personal space: Differentially private cybersickness detection in immersive virtual reality environments. In *2024 IEEE International Symposium on Mixed and Augmented Reality (ISMAR)*, pp. 11–20. IEEE, 2024. 1, 2, 3, 4, 5, 6, 9

[36] R. K. Kundu, R. Islam, J. Quarles, and K. A. Hoque. Litevr: Interpretable and lightweight cybersickness detection using explainable ai. In *2023 IEEE Conference Virtual Reality and 3D User Interfaces (VR)*, pp. 609–619. IEEE, 2023. 2, 3, 6

[37] T. M. Lee, J.-C. Yoon, and I.-K. Lee. Motion sickness prediction in stereoscopic videos using 3d convolutional neural networks. *IEEE transactions on visualization and computer graphics*, 25(5):1919–1927, 2019. 2

[38] J. Li, A. R. Chowdhury, K. Fawaz, and Y. Kim. {Kalɛido}:{Real-Time} privacy control for {Eye-Tracking} systems. In *30th USENIX security symposium (USENIX security 21)*, pp. 1793–1810, 2021. 1, 2, 9

[39] L. Li, F. Yu, D. Shi, J. Shi, Z. Tian, J. Yang, X. Wang, and Q. Jiang. Application of virtual reality technology in clinical medicine. *American journal of translational research*, 9(9):3867, 2017. 1

[40] W.-F. Lin, D.-Y. Tsai, L. Tang, C.-T. Hsieh, C.-Y. Chou, P.-H. Chang, and L. Hsu. Onnc: A compilation framework connecting onnx to proprietary deep learning accelerators. In *2019 IEEE International Conference on Artificial Intelligence Circuits and Systems (AICAS)*, pp. 214–218. IEEE, 2019. 7

[41] A. Liu, L. Xia, A. Duchowski, R. Bailey, K. Holmqvist, and E. Jain.

Differential privacy for eye-tracking data. In *Proceedings of the 11th ACM Symposium on Eye Tracking Research & Applications*, pp. 1–10, 2019. 2

[42] S. M. Lundberg and S.-I. Lee. A unified approach to interpreting model predictions. *Advances in neural information processing systems*, 30, 2017. 3, 4

[43] T. Luong and C. Holz. Characterizing physiological responses to fear, frustration, and insight in virtual reality. *IEEE Transactions on Visualization and Computer Graphics*, 28(11):3917–3927, 2022. 2

[44] L. Melis, C. Song, E. De Cristofaro, and V. Shmatikov. Exploiting unintended feature leakage in collaborative learning. In *2019 IEEE symposium on security and privacy (SP)*, pp. 691–706. IEEE, 2019. 1

[45] MelonLoader. Melonloader community. https://melonwiki.xyz/#/, 2022. Accessed: March 28, 2025. 7

[46] Y. Meng, Y. Zhan, J. Li, S. Du, H. Zhu, and X. Shen. De-anonymizing avatars in virtual reality: Attacks and countermeasures. *IEEE Transactions on Mobile Computing*, 2024. 2

[47] M. R. Miller, V. Nair, E. Han, C. DeVeaux, C. Rack, R. Wang, B. Huang, M. E. Latoschik, J. F. O'Brien, and J. N. Bailenson. Effect of duration and delay on the identifiability of vr motion. In *2024 IEEE 25th International Symposium on a World of Wireless, Mobile and Multimedia Networks (WoWMoM)*, pp. 70–75. IEEE, 2024. 2

[48] R. Miller, N. K. Banerjee, and S. Banerjee. Combining real-world constraints on user behavior with deep neural networks for virtual reality (vr) biometrics. In *2022 IEEE Conference on Virtual Reality and 3D User Interfaces (VR)*, pp. 409–418. IEEE, 2022. 2

[49] I. Mironov. Rényi differential privacy. In *2017 IEEE 30th computer security foundations symposium (CSF)*, pp. 263–275. IEEE, 2017. 4

[50] A. G. Moore, T. D. Do, N. Ruozzi, and R. P. McMahan. Identifying virtual reality users across domain-specific tasks: A systematic investigation of tracked features for assembly. In *2023 IEEE International Symposium on Mixed and Augmented Reality (ISMAR)*, pp. 396–404. IEEE, 2023. 2

[51] A. G. Moore, R. P. McMahan, H. Dong, and N. Ruozzi. Personal identifiability and obfuscation of user tracking data from vr training sessions. In *2021 IEEE International Symposium on Mixed and Augmented Reality (ISMAR)*, pp. 221–228. IEEE, 2021. 2

[52] D. Munoz, J. Broughton, J. Goldring, and I. Armstrong. Age-related performance of human subjects on saccadic eye movement tasks. *Experimental brain research*, 121:391–400, 1998. 2

[53] G. Muthukrishnan and S. Kalyani. Grafting laplace and gaussian distributions: A new noise mechanism for differential privacy. *IEEE Transactions on Information Forensics and Security*, 2023. 4

[54] V. Nair, W. Guo, J. Mattern, R. Wang, J. F. O'Brien, L. Rosenberg, and D. Song. Unique identification of 50,000+ virtual reality users from head & hand motion data. In *32nd USENIX Security Symposium (USENIX Security 23)*, pp. 895–910, 2023. 1, 2, 3, 9

[55] V. Nair, W. Guo, J. F. O'Brien, L. Rosenberg, and D. Song. Deep motion masking for secure, usable, and scalable real-time anonymization of ecological virtual reality motion data. In *2024 IEEE Conference on Virtual Reality and 3D User Interfaces Abstracts and Workshops (VRW)*, pp. 493–500. IEEE, 2024. 2, 3

[56] V. C. Nair, G. Munilla-Garrido, and D. Song. Going incognito in the metaverse: Achieving theoretically optimal privacy-usability tradeoffs in vr. In *Proceedings of the 36th Annual ACM Symposium on User Interface Software and Technology*, pp. 1–16, 2023. 2, 6, 9

[57] M. Nasr, R. Shokri, et al. Improving deep learning with differential privacy using gradient encoding and denoising. *arXiv preprint arXiv:2007.11524*, 2020. 4

[58] M.-I. Nicolae, M. Sinn, M. N. Tran, B. Buesser, A. Rawat, M. Wistuba, V. Zantedeschi, N. Baracaldo, B. Chen, H. Ludwig, et al. Adversarial robustness toolbox v1. 0. 0. *arXiv preprint arXiv:1807.01069*, 2018. 4

[59] I. Olade, C. Fleming, and H.-N. Liang. Biomove: Biometric user identification from human kinesiological movements for virtual reality systems. *Sensors*, 20(10):2944, 2020. 2

[60] S. Özdel, Y. Rong, B. M. Albaba, Y.-L. Kuo, X. Wang, and E. Kasneci. A transformer-based model for the prediction of human gaze behavior on videos. In *Proceedings of the 2024 Symposium on Eye Tracking Research and Applications*, pp. 1–6, 2024. 3

[61] A. Paszke, S. Gross, F. Massa, A. Lerer, J. Bradbury, G. Chanan, T. Killeen, Z. Lin, N. Gimelshein, L. Antiga, et al. Pytorch: An imperative style, high-performance deep learning library. *Advances in neural information processing systems*, 32, 2019. 4

[62] D. Reilly and L. Fan. A comparative evaluation of differentially private image obfuscation. In *2021 Third IEEE International Conference on Trust, Privacy and Security in Intelligent Systems and Applications (TPS-ISA)*, pp. 80–89. IEEE, 2021. 1

[63] P. Reyero Lobo and P. Perez. Heart rate variability for non-intrusive cybersickness detection. In *ACM International Conference on Interactive Media Experiences*, pp. 221–228, 2022. 2

[64] J. N. Setu, J. M. Le, R. K. Kundu, B. Giesbrecht, T. Höllerer, K. A. Hoque, K. Desai, and J. Quarles. Predicting and explaining cognitive load, attention, and working memory in virtual multitasking. *IEEE Transactions on Visualization and Computer Graphics*, 2025. 2, 3

[65] R. Shokri, M. Stronati, C. Song, and V. Shmatikov. Membership inference attacks against machine learning models. In *2017 IEEE symposium on security and privacy (SP)*, pp. 3–18. IEEE, 2017. 1

[66] C. Slocum, Y. Zhang, N. Abu-Ghazaleh, and J. Chen. Going through the motions:{AR/VR} keylogging from user head motions. In *32nd USENIX Security Symposium (USENIX Security 23)*, pp. 159–174, 2023. 9

[67] C. Song, T. Ristenpart, and V. Shmatikov. Machine learning models that remember too much. In *Proceedings of the 2017 ACM SIGSAC Conference on computer and communications security*, pp. 587–601, 2017. 1

[68] L. Tabbaa, R. Searle, S. M. Bafti, M. M. Hossain, J. Intarasisrisawat, M. Glancy, and C. S. Ang. Vreed: Virtual reality emotion recognition dataset using eye tracking & physiological measures. *Proceedings of the ACM on interactive, mobile, wearable and ubiquitous technologies*, 5(4):1–20, 2021. 1, 2, 4

[69] A. Vaswani, N. Shazeer, N. Parmar, J. Uszkoreit, L. Jones, A. N. Gomez, Ł. Kaiser, and I. Polosukhin. Attention is all you need. *Advances in neural information processing systems*, 30, 2017. 3

[70] Q. J. Wang and R. P. McMahan. Gender identification of vr users by machine learning tracking data. In *2024 IEEE Conference on Virtual Reality and 3D User Interfaces Abstracts and Workshops (VRW)*, pp. 827–828. IEEE, 2024. 2

[71] Q. J. Wang, A. G. Moore, N. N. Chawla, and R. P. McMahan. Cross-domain gender identification using vr tracking data. In *2024 IEEE International Symposium on Mixed and Augmented Reality (ISMAR)*, pp. 180–189. IEEE, 2024. 2

[72] Z. Wang and F. Lu. Tasks reflected in the eyes: Egocentric gaze-aware visual task type recognition in virtual reality. *IEEE Transactions on Visualization and Computer Graphics*, 2024. 2

[73] E. Wilson, A. Ibragimov, M. J. Proulx, S. D. Tetali, K. Butler, and E. Jain. Privacy-preserving gaze data streaming in immersive interactive virtual reality: Robustness and user experience. *arXiv preprint arXiv:2402.07687*, 2024. 2, 3

[74] E. Wood, T. Baltrušaitis, L.-P. Morency, P. Robinson, and A. Bulling. Gazedirector: Fully articulated eye gaze redirection in video. In *Computer Graphics Forum*, vol. 37, pp. 217–225. Wiley Online Library, 2018. 2

[75] Z. Yang, Z. Sarwar, I. Hwang, R. Bhaskar, B. Y. Zhao, and H. Zheng. Can virtual reality protect users from keystroke inference attacks? In *33rd USENIX Security Symposium (USENIX Security 24)*, pp. 2725–2742, 2024. 2

[76] A. Yousefpour, I. Shilov, A. Sablayrolles, D. Testuggine, K. Prasad, M. Malek, J. Nguyen, S. Ghosh, A. Bharadwaj, J. Zhao, et al. Opacus: User-friendly differential privacy library in pytorch. *arXiv preprint arXiv:2109.12298*, 2021. 5

[77] R. Zemblys, D. C. Niehorster, O. Komogortsev, and K. Holmqvist. Using machine learning to detect events in eye-tracking data. *Behavior research methods*, 50:160–181, 2018. 2

[78] X. Zhou, I. Viola, S. Rossi, and P. Cesar. Comparison of visual saliency for dynamic point clouds: Task-free vs. task-dependent. *IEEE Transactions on Visualization and Computer Graphics*, 2025. 2